\documentclass[
    ,final            
  ]
  {aipproc}

\layoutstyle{6x9}

\begin{document}

\title{Evidence for a New Light Boson from Cosmological Gamma-Ray Propagation?}
\classification{14.80.Mz, 95.30.-k, 95.85.Pw, 95.85.Ry, 98.70.Rz. 98.70.Vc, 98.70.Sa}
\keywords{cosmic rays, gamma rays, axion, photon propagation}


\author{Marco Roncadelli}{address={INFN, Sezione di Pavia, via A. Bassi 6, I -- 27100 Pavia, Italy}}

\author{Alessandro De Angelis}{address={Dipartimento di Fisica, Universit\`a di Udine, Via delle Scienze 208, I -- 33100 Udine, and INAF and INFN, Sezioni di Trieste, Italy},altaddress={also at IST, Lisboa, Portugal}}

\author{Oriana Mansutti}{address={Dipartimento di Fisica, Universit\`a di Udine, Via delle Scienze 208, I -- 33100 Udine, and INFN, Sezione di Trieste, Italy}}

\begin{abstract}

An anomalously large transparency of the Universe to gamma rays has recently been discovered by the Imaging Atmospheric Cherenkov Telescopes (IACTs) H.E.S.S. and MAGIC. We show that observations can be reconciled with standard blazar emission models provided photon oscillations into a very light Axion-Like Particle occur in extragalactic magnetic fields. A quantitative estimate of this effect is successfully applied to the blazar 3C\,279. Our prediction can be tested with the satellite-borne {\it Fermi}/LAT detector as well as with the ground-based IACTs  H.E.S.S., MAGIC, CANGAROO\,III, VERITAS and the Extensive Air Shower arrays ARGO-YBJ and MILAGRO. Our result also offers an important observational test for models of dark energy wherein quintessence is coupled to the photon through an effective dimension-five operator\footnote{Proceeding of the Conference ``Frontiers of Fundamental and Computational Physics'', AIP Conference Proceedings 1018 (2008).}.
\end{abstract}

\maketitle


\section{Introduction}

An impressive amount of information about the Universe in the very-high-energy (VHE) band has been collected over the last few years by the Imaging Atmospheric Cherenkov Telescopes (IACTs) H.E.S.S., MAGIC, CANGAROO III and VERITAS. Observations carried out by these IACTs concern gamma-ray sources over an extremely wide interval of distances, ranging from the parsec scale for Galactic objects up to the Gigaparsec scale for the farthest detected blazar 3C279. This circumstance allows not only to infer the intrinsic properties of the sources, but also to probe the nature of photon propagation throughout cosmological distances. 

The latter isssue becomes particularly important in the energy range above $100 \, {\rm GeV}$, where the horizon of the observable Universe rapidly shrinks as the energy further increases. This comes about because photons from distant sources scatter off background photons permeating the Universe, thereby disappearing into electron-positron pairs~\cite{stecker1971}. The corresponding cross section $\sigma (\gamma \gamma \to e^+ e^-)$ turns out to peak when the VHE photon energy $E$ and the background photon energy $\epsilon$ are related by $\epsilon \simeq (500 \, {\rm GeV}/E) \, {\rm eV}$, so that the resulting cosmic opacity is dominated by the interaction with ultraviolet/optical/infrared photons of the diffuse extragalactic background -- usually called {\it extragalactic background light} (EBL) -- for observations performed by IACTs. 

Owing to the absorption process in question, photon propagation is controlled by the optical depth ${\tau}(E,D)$, with $D$ denoting the source distance. Therefore, the observed photon spectrum $\Phi_{\rm obs}(E,D)$ is related to the emitted one $\Phi_{\rm em}(E)$ by 
\begin{equation}
\label{a0}
\Phi_{\rm obs}(E,D) = e^{- \tau(E,D)} \ \Phi_{\rm em}(E)~. 
\end{equation}
Unlike the CMB, the EBL is produced by galaxies during the whole age of the Universe and possibly by a first generation of stars formed before galaxies were assembled. Based on stellar evolution models in galaxies as well as on deep galaxy counts, several groups have attempted a determination of the spectral energy distribution of the EBL and ultimately of the optical depth ${\tau}(E,D)$ for $100~{\rm GeV} < E < 100 \, {\rm TeV}$~\cite{kneiske}. Because galaxies were brigther in the past than they are now, evolutionary effects should in principle be included in the evaluation of ${\tau}(E,D)$ but they become unimportant at sufficiently low redshift and will therefore be neglected throughout. Correspondingly we have ${\tau}(E,D) \simeq D/{\lambda}_{\gamma}(E)$, with ${\lambda}_{\gamma}(E)$ denoting the photon mean free path for $\gamma \gamma \to e^+ e^-$. As a consequence, Eq. (\ref{a0}) becomes
\begin{equation}
\label{a1}
\Phi_{\rm obs}(E,D) \simeq e^{- D/{\lambda}_{\gamma}(E)} \ \Phi_{\rm em}(E)~.
\end{equation}
The function ${\lambda}_{\gamma}(E)$ has been computed within realistic models of the EBL and is reported in Fig. 1 (from ref.~\cite{CoppiAharonian}). 

\begin{figure}
\centering
\includegraphics[width=.65\textwidth]{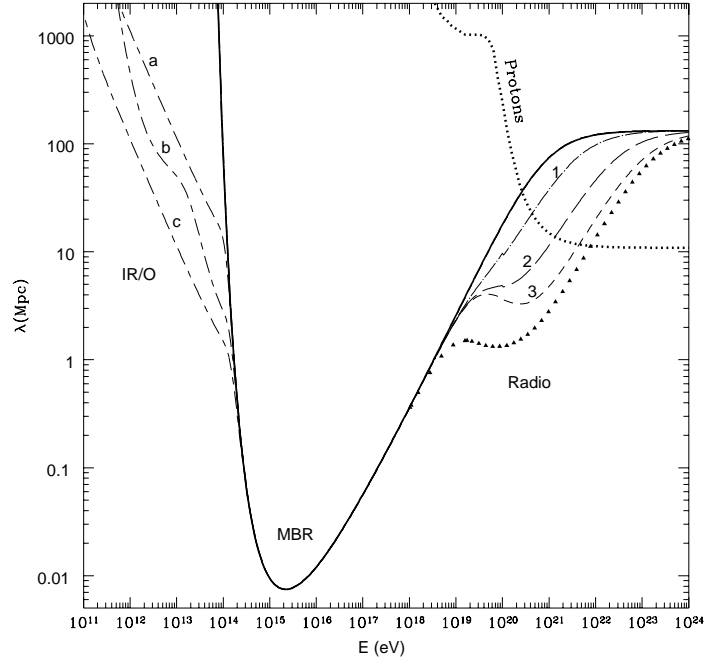}
\caption{\label{CoppiAharonian}
	Pair-production photon mean free path as a function of beam photon energy (from ref.~\cite{CoppiAharonian}).}
\end{figure}

We see that ${\lambda}_{\gamma}(E)$ decreases roughly like a power law from the Hubble radius $4.3 \, {\rm Gpc}$ slightly below $100 \, {\rm GeV}$ to about $1 \, {\rm Mpc}$ at $100 \, {\rm TeV}$. Thus, Eq. (\ref{a1}) entails that the observed flux is {\it exponentially} suppressed both at high energy and at large distances, so that sufficiently far-away sources become {\it hardly visible} in the VHE range. Moreover, the observed spectrum of distant sources gets {\it much steeper} than the emitted one.

Yet, observations have {\it not} detected the behaviour predicted by Eq. (\ref{a1}). A first indication in this direction was reported by the H.E.S.S. collaboration in connection with the discovery of the two blazars H2356-309 ($z = 0.165$) and 1ES1101-232 ($z = 0.186$) at $E \sim 1 \, {\rm TeV}$~\cite{aharonian:nature06}. Stronger evidence comes from the observation of the blazar 3C\,279 ($z = 0.536$) at $E \sim 0.5 \, {\rm TeV}$ by the MAGIC collaboration~\cite{3c}. In particular, the signal from 3C279 collected by MAGIC in the region $E<220$ GeV has more or less the same statistical significance as the one in the range 220 GeV $< E <$ 600 GeV ($6.1 \sigma$ in the former case, $5.1 \sigma$ in the latter)~\cite{3c}. 

Turning the argument around and assuming {\it standard} photon propagation, the observed spectrum can only be reproduced by an emission spectrum {\it much harder} than for any other observed blazar. Such a hard spectrum is also largely inconsistent with the predictions of current blazar models~\cite{fossati}. So, we are led to believe that the effect under consideration ought to be due to an {\it anomalous} photon propagation over cosmological distances rather than to new phenomena occurring inside the blazars themselves. From this point of view, the Universe appears to be {\it more transparent} to VHE gamma rays than previously thought, up to the point that a conflict with the standard scenario can be envisaged.
 
A cautionary remark is in order. Available observational information is insufficient to fully settle the issue and in fact it has been pointed out that an anomalously hard emission spectrum can be produced within unconventional blazar models~\cite{steckerscully}. However, according to this option it looks difficult to explain why the physical mechanisms occurring in the most distant blazars are so different from those described by standard blazar models. 

A way out of the considered difficulty has recently been proposed by the present authors and rests upon an oscillation mechanism occurring in extragalactic magnetic fields, whereby a photon can become a new very light spin-zero particle, named {\it Axion-Like Particle} (ALP)~\cite{drm}. Once produced, the ALP travels unimpeded throughout the Universe and can convert back to a photon before reaching the Earth, thereby acting as if the observed photons had travelled a distance {\it largely in excess} of their mean free path. Thanks to the exponential dependence of the observed flux on 
${\lambda}_{\gamma}(E)$, even a small increase in ${\lambda}_{\gamma}(E)$ gives rise to a large effect. More specifically, this mechanism yields an observed spectrum {\it much harder} than the one predicted by Eq. (\ref{a1}) for distant blazars, thereby leading to agreement with observations for {\it standard} emission spectra.

Our aim is to review the main features of our proposal as well as its application to blazar 3C\,279.

\section{Photon-ALP oscillations}

\subsection{Motivations for ALPs}

In spite of the enormous success scored by the Standard Model of strong, weak and electromagnetic interactions in describing physics at energies up to the Fermi scale $G_F^{- 1/2} \simeq 250 \, {\rm GeV}$, nobody would seriously regard it as the final theory. Instead, on the basis of phenomenological as well as conceptual reasons, the Standard Model is currently viewed as the low-energy manifestation of some more fundamental and richer theory of all elementary-particle interactions including gravity. Therefore, the lagrangian of the Standard Model is expected to be modified by small non-renormalizable terms describing interactions among known and new particles. 

Attempts to go beyond the Standard Model are the central research topic in high-energy physics since almost three decades and each specific proposal is characterized by a set of new particles along with their mass spectrum and interactions with the standard world. ALPs turn out to be a generic prediction of many extensions of the Standard Model and have attracted considerable interest over the last few years. Besides than in four-dimensional 
models~\cite{masso1}, they naturally arise in the context of compactified Kaluza-Klein theories~\cite{kk} as well as in superstring theories~\cite{superstring}. Moreover, it has been argued that an ALP with mass $m \sim 10^{-33} \, {\rm eV}$ is a good candidate for quintessential dark energy~\cite{carroll} which presumably triggers the present accelerated expansion of the Universe.

Specifically, ALPs are spin-zero light bosons defined by the following low-energy effective lagrangian
\begin{equation}
\label{a1a}
{\cal L}_{\rm ALP} \ = \ 
\frac{1}{2} \, \partial^{\mu} \, a \, \partial_{\mu} \, a - \frac{1}{2} 
\, m^2 \, a^2 - \frac{1}{4 M} \, F^{\mu \nu} \, \tilde F_{\mu \nu} \, a~,
\end{equation}
where $F^{\mu \nu}$ is the electromagnetic field strength, $\tilde F_{\mu \nu}$ is its dual and $a$ denotes the ALP field. According to the above view, it is assumed that the inverse two-photon coupling $M$ is much larger than $G_F^{- 1/2} \simeq 250 \, {\rm GeV}$. On the other hand, the ALP mass $m$ is supposed to be much smaller than $G_F^{- 1/2} \simeq 250 \, {\rm GeV}$ and for definiteness we take $m < 1 \, {\rm eV}$. As far as generic ALPs are concerned, the parameters $M$ and $m$ are regarded as {\it independent}. The situation is at variance with the case of the standard {\it Axion}~\cite{assione}, for which 
the relationship
\begin{equation}
\label{a2}
m = 0.7 \, k \cdot \left( \frac{10^{10} \, {\rm GeV}}{M} \right) \, {\rm eV}
\end{equation}
holds, with $k$ a model-dependent constant of order one~\cite{cgn}.

\subsection{General considerations}

A characteristic feature of ALPs is the trilinear $\gamma$-$\gamma$-$a$ vertex described by the last term in 
${\cal L}_{\rm ALP}$, whereby one ALP couples to two photons. This interaction gives rise to photon-ALP conversion, which leads in turn to a very interesting variety of physical processes, including the one to which the present paper is devoted. 

One of these processes consists in ALP photo-production through the {\it Primakoff process}, which takes place when an incoming photon scatters on a charged particle thereby becoming an ALP upon the exchange of a virtual photon. Hot, dense plasmas in stellar cores are ideal environments where the Primakoff process involving thermal photons can occur. Once produced, the ALPs escape because their mean free path is much larger than the stellar radius, thereby carrying off energy. Owing to the virial equilibrium, the core has a negative specific heat. Therefore it reacts to such an energy loss by getting hotter. As a result, the rate of nuclear reactions sharply increases, bringing about a substantial change in the observed properties of stars. Still, current models of stellar evolution are in fairly good agreement with observations. Hence, $M$ has to be large enough to provide a sufficient suppression of unwanted ALP effects. This argument has been applied in a quantitative fashion to the Sun, to main-sequence stars and to red-giants stars, with the result~\cite{Raffelt1990} 
\begin{equation}
\label{a3}
M > 10^{10} \, {\rm GeV}~. 
\end{equation} 
Remarkably enough, for $m < 0.02 \, {\rm eV}$ almost the same conclusion, namely
\begin{equation}
\label{a4}
M > 0.86 \cdot 10^{10} \, {\rm GeV}~,
\end{equation}
has been reached by the negative result of the CAST experiment designed to detect ALPs emitted by the Sun~\cite{cast}.

Another process implied by the $\gamma$-$\gamma$-$a$ vertex is {\it direct} photon-ALP conversion without the need of charged particles playing the role of catalysts, provided an {\it external} magnetic (electric) field is present. Whenever the external field extends over a large region and the momentum transfer is small, photon-ALP conversion becomes coherent and can be regarded as an {\it oscillation} phemomenon. Indeed, the $\gamma$-$\gamma$-$a$ vertex produces an {\it off-diagonal} element in the mass matrix for the photon-ALP system in the presence of an external field. Therefore, the interaction eigenstates {\it differ} from the propagation eigenstates and photon-ALP oscillations show 
up~\cite{RaffeltStodolsky}. The situation is analogous to what happens in the case of massive neutrinos with different flavours, apart from an important difference. All neutrinos have equal spin, and so neutrino oscillations can freely occur. Instead, ALPs are supposed to have spin zero whereas the photon has spin one, and so one of them can transform into the other only if the spin mismatch is compensated for by an external field. 

One consequence of photon-ALP oscillations is that a lower bound on $M$ stronger than condition (\ref{a3}) can be derived for ALPs with mass $m < 10^{- 10} \, {\rm eV}$. In this connections, two methods have been put forward. One concerns the energetics of the supernova 1987a. Because the emitted neutrinos have been observed and the whole energy budget is rather well known, an upper bound on the ALP flux can be derived~\cite{raffeltmasso}. Alternatively, observation of time-lag between opposite-polarization modes in pulsar radio emission similarly yields an upper bound on the two-photon coupling of an ALP~\cite{mohanti}. In either case, one gets
\begin{equation}
\label{a5}
M > 3 \cdot 10^{11} \, {\rm GeV}~.
\end{equation}

\subsection{Application to VHE gamma-ray observations}

Our proposal now starts to become clear. We imagine that photons emitted by a distant blazar can oscillate into ALPs in the presence of cosmic magnetic fields. So, the emitted flux gets reduced along the line-of-sight because some photons become ALPs. If this were the whole story, the observable prediction would merely be a {\it dimming}~\cite{dimming}. Things can be totally {\it different} when photon absorption becomes important. To see how this comes about, suppose that a sizeable fraction of the emitted photons convert into ALPs close enough to the source and that a nonnegligible fraction of the ALPs in question are in turn converted back into photons close enough to the Earth. Because ALPs propagate unimpeded, in such a situation the observed photon flux can be considerably {\it larger} than the one predicted by Eq. (\ref{a1}). Below, we will turn this qualitative picture into a quantitative estimate.

In principle, we have to evaluate the probability $P_{\gamma \to \gamma}(D)$ that a photon remains a photon after propagation from the source to us when allowance is made for photon-ALP oscillation as well as for photon absorption. 

However, an exact treatment would be exceedingly difficult, mainly because of the uncertainty concerning the configuration of the magnetic field responsible for photon-ALP oscillations. Actually, the line-of-sight to a distant blazar is expected to cross magnetic fields on different scales. A magnetic field is likely present in the source itself. In addition, the Galactic magnetic field can give a nontrivial contribution to the effect under consideration. Finally, extragalactic magnetic fields can play an important role. Throughout, we restrict our attention to extragalactic magnetic fields, whose existence has been demonstrated very recently by AUGER observations~\cite{auger}. A complementary picture involving only magnetic fields in the source and in the Milky Way has been considered in ref.~\cite{simet}. 

Unfortunately, almost nothing is known about the morphology of extragalactic magnetic fields, which reflects both their cosmic origin and the evolutionary history of baryonic matter. While it is evident that their coherence length cannot be arbitrarily large, no reliable estimate of its value is presently available. As far as our analysis is concerned, this means that we cannot suppose that extragalactic magnetic fields are constants over the whole distance to the source -- yet, their spatial dependence is unknown. The usual way out of this difficulty amounts to suppose that extragalactic magnetic fields ${\bf B}$ have a domain-like structure~\cite{ckpt}. That is, ${\bf B}$ is assumed to be constant over a domain of size $L_{\rm dom}$ equal to its coherence length, with ${\bf B}$ randomly changing its direction from one domain to another but keeping approximately the same strength. Reference values to be used throughout are \mbox{$B \simeq 10^{- 9} \, {\rm G}$} and \mbox{$L_{\rm dom} \simeq 1 \, {\rm Mpc}$}, which are close to existing upper limits but consistent with them~\cite{fur}. Remarkably enough, such a picture of cosmic magnetic fields turns out to be consistent with AUGER observations~\cite{dpr}.

Thus, the whole propagation process of the photon beam can be recovered by iterating the propagation over a single domain as many times as the number of domains crossed by the beam, taking each time a random value for the angle $\theta$ 
between ${\bf B}$ and a fixed fiducial direction. In this way, we are effectively led to the much easier problem of photon-ALP oscillation in a {\it constant} magnetic field. 

Another simplification is brought about by the fact that photon absorption is insensitive to the specific properties of the photon-ALP oscillation -- it only depends on the photon mean free path ${\lambda}_{\gamma}$. So, we can first identify the regime in which $P_{\gamma \to \gamma}$ is maximal over a single magnetic domain and next we can work out how much this probability is suppressed by photon absorption. 

Consider in the first place the propagation of a photon beam over one domain without photon absorption. Since now ${\bf B}$ is constant, the probability that a photon converts to an ALP after a distance $y$ can be computed exactly and reads~\cite{RaffeltStodolsky}
\begin{equation}
\label{a16}
P_{\gamma \to a}^{(0)}(y) = {\rm sin}^2 2 \alpha \  {\rm sin}^2
\left( \frac{\Delta_{\rm osc} \, y}{2} \right)~,
\end{equation}
where the photon-ALP mixing angle $\alpha$ is
\begin{equation}
\label{a16m}
\alpha = \frac{1}{2} \, {\rm arcsin} \left( \frac{B}{M \, {\Delta}_{\rm osc}} \right)
\end{equation}
and the oscillation wave number reads
\begin{equation}
\label{a17}
{\Delta}_{\rm osc} = 
\left[\left( \frac{m^2 - {\omega}_{\rm pl}^2}{2 E} \right)^2 + 
\left( \frac{B}{M} \right)^2 \right]^{1/2}~,
\end{equation}
so that the  oscillation length is  \mbox{$L_{\rm osc} = 2 \pi / {\Delta}_{\rm osc}$}. Actually, eq.~(\ref{a17}) pertains to the situation in which the beam propagates in a cold intergalactic plasma with plasma frequency 
\begin{equation}
\label{a6}
{\omega}_{\rm pl} = \left( \frac{4 \pi \alpha n_e}{m_e} \right)^{1/2} \simeq 3.69 \cdot 10^{- 11} \, \left( 
\frac{n_e}{{\rm cm}^{- 3}} \right)^{1/2} \, {\rm eV}~, 
\end{equation}
where~$n_e$ denotes the electron density. Because we are dealing with weak magnetic fields, their contribution to the vacuum refractive index is totally negligible.

A look back at Eqs. (\ref{a16}), (\ref{a16m}) and (\ref{a17}) shows that the photon-ALP transition probability is maximal in the strong-mixing regime, namely for \mbox{$E \gg |m^2 - {\omega}^2_{\rm pl}|M/2 B$}. A conservative estimate of the density of intergalactic plasma yields \mbox{$n_e \simeq 10^{-7} \, {\rm cm}^{-3}$}~\cite{peebles}, resulting in the plasma frequency \mbox{${\omega}_{\rm pl} \simeq 1.17 \cdot 10^{-14} \, {\rm eV}$}. Therefore, the strong-mixing condition takes the explicit form $|(m/10^{- 10} \, {\rm eV})^2 - 1.37 \cdot 10^{- 8}| \ll 0.38 (E/{\rm GeV}) ( B/10^{-9} \, {\rm G})$ $(10^{10} \, {\rm GeV}/M)$. Since we are interested in the energy range $E > 10^2 \, {\rm GeV}$, we find that for e.g.\ \mbox{$M > 4 \cdot 10^{11} \, {\rm GeV}$} the bound \mbox{$m \ll 10^{- 10} \, {\rm eV}$} has to be satisfied. Recalling Eq. (\ref{a2}), we see that the standard axion is excluded. We also remark that the present mechanism works for arbitrarily small values of $m$, 
provided $M$ happens to be considerably smaller than the Planck mass $M_P \simeq 1.22 \cdot 10^{19} \, {\rm GeV}$. As a consequence, our result also applies to models of dark energy wherein quintessence enjoys a photon coupling described by ${\cal L}_{\rm ALP}$~\cite{carroll}, thus ultimately providing an important observational test for these models.

\section{Energy spectrum for blazar 3C\,279}

It is straightforward to solve the beam propagation equation once photon absorption by the EBL is taken into account and produces a finite photon mean free path ${\lambda}_{\gamma}$~\cite{ckpt}. In the strong-mixing regime, the probability for a photon to become an ALP after a distance $y \leq L_{\rm dom}$ becomes~
\begin{equation}
\label{hm2} 
P_{\gamma \to a}(y) \simeq \frac{1}{2} \, e^{- y/(2 {\lambda}_{\gamma})} \, {\rm sin}^2 \left( \frac{y \delta}{2 
{\lambda}_{\gamma}} \right)~,
\end{equation}
whereas the probability that a photon remains a photon now reads
\begin{equation}
\label{hm3} 
P_{\gamma \to \gamma}(y) \simeq \frac{1}{2} \, e^{- y/{\lambda}_{\gamma}} \, \left[ 1 
+ {\rm cos}^2 \left( \frac{y \delta}{2 {\lambda}_{\gamma}} \right) \right]~,
\end{equation}
where we have introduced the dimensionless parameter
\begin{equation}
\label{as1}
\delta \equiv \frac{B \, {\lambda}_{\gamma} }{M} \simeq 0.11 \left( \frac{B}{10^{-9}\, {\rm G}} \right) \left( \frac{10^{11} \, {\rm GeV}}{M} \right)
\left( \frac{{\lambda}_{\gamma}}{{\rm Mpc}} \right)~.
\end{equation}
The relation between the photon energy $E$ and the source redshift $z$ yields for 3C\,279 ${\lambda}_{\gamma} \simeq 450$~Mpc at $E = 500 \, {\rm GeV}$~\cite{figuraOriz}, so that we have $\delta \simeq 12.4$ in this case study.

Over distances \mbox{$y \gg L_{\rm dom}$}, the transition probabilities $P_{\gamma \to a}(y)$ and $P_{\gamma \to \gamma}(y)$ arise as the incoherent average of Eqs.~(\ref{hm2}) and (\ref{hm3}) over \mbox{$N \simeq (y/ L_{\rm dom})$} domains crossed by the beam, respectively.
Assuming (as before) that the beam propagates along the $y$ direction and choosing the $x$ and 
$z$ directions arbitrarily in the orthogonal plane, the problem becomes truly three-dimensional, because of the random orientation of the magnetic field.
Consequently, the beam state is described by the vector $({\gamma}_x, {\gamma}_z, a)$.

We have derived the propagation equations describing the absorption of photons due to the interaction with the EBL and their oscillations to ALPs (and vice-versa). As in Ref.~\cite{ckpt}, we are led to the transfer equation
\begin{equation}
\left(\!
\begin{array}{c} 
\gamma_x \\ \gamma_z \\ a
\end{array} 
\!\!\right)
=
{\rm e}^{i\,E\,y} \,
\left[ \, T_0 \, {\rm e}^{\lambda_0\,y}
+T_1 \, {\rm e}^{\lambda_1\,y} + T_2 \, {\rm e}^{\lambda_2\,y} \, 
\right]\!\!
\left(\!
\begin{array}{c}
\gamma_x \\ \gamma_z \\ a                      \label{eq:evolution}
\end{array}
\!\!\right)_{\!\!\!0}   \!
\end{equation}

\phantom{where}

\noindent
where
\begin{eqnarray}
\lambda_0 &\equiv& -\,\frac{1}{2\,{\lambda}_{\gamma}} \,\,,
\nonumber
\\
\lambda_1 &\equiv& - \, \frac{1}{4\,{\lambda}_{\gamma}} \, \left[ 1 +
\sqrt{1-4\,\delta^2} \right] \,\,,
\\
\lambda_2 &\equiv& - \, \frac{1}{4\,{\lambda}_{\gamma}} \, \left[ 1 -
\sqrt{1-4\,\delta^2} \right] \,\,,
\nonumber 
\end{eqnarray}

\begin{eqnarray}
T_0 &\equiv& \left( \begin{array}{ccc}
{\rm sin}^2 \theta & -\, {\rm cos} \theta \, {\rm sin} \theta & 0 \\
-\, {\rm cos} \theta \, {\rm sin} \theta & {\rm cos}^2 \theta & 0 \\
0 & 0 & 0
\end{array} \right) \,\,, 
\nonumber
\\
T_1 &\equiv& \left( \begin{array}{ccc}
\frac{1+\sqrt{1-4\,\delta^2}}{2\,\sqrt{1-4\,\delta^2}}
\, {\rm cos}^2 \theta &
\frac{1+\sqrt{1-4\,\delta^2}}{2\,\sqrt{1-4\,\delta^2}}
\, {\rm cos} \theta \, {\rm sin} \theta &
-\,\frac{\delta}{\sqrt{1-4\,\delta^2}} \,{\rm cos} \theta \\
 \frac{1+\sqrt{1-4\,\delta^2}}{2\,\sqrt{1-4\,\delta^2}}
 \, {\rm cos} \theta \, {\rm sin} \theta &
\frac{1+\sqrt{1-4\,\delta^2}}{2\,\sqrt{1-4\,\delta^2}} \, {\rm sin}^2 \theta &
-\,\frac{\delta}{\sqrt{1-4\,\delta^2}} \,{\rm sin} \theta \\
\frac{\delta}{\sqrt{1-4\,\delta^2}} \,{\rm cos} \theta &
\frac{\delta}{\sqrt{1-4\,\delta^2}} \,{\rm sin} \theta &
-\,\frac{1-\sqrt{1-4\,\delta^2}}{2\,\sqrt{1-4\,\delta^2}}
\end{array} \right) \,\,, 
\\
T_2 &\equiv& \left( \begin{array}{ccc}
-\,\frac{1-\sqrt{1-4\,\delta^2}}{2\,\sqrt{1-4\,\delta^2}} \, {\rm cos}^2 \theta &
-\,\frac{1-\sqrt{1-4\,\delta^2}}{2\,\sqrt{1-4\,\delta^2}}
\, {\rm cos} \theta \, {\rm sin} \theta &
\frac{\delta}{\sqrt{1-4\,\delta^2}} \,{\rm cos} \theta \\
-\,\frac{1-\sqrt{1-4\,\delta^2}}{2\,\sqrt{1-4\,\delta^2}}
\, {\rm cos} \theta \, {\rm sin} \theta &
-\,\frac{1-\sqrt{1-4\,\delta^2}}{2\,\sqrt{1-4\,\delta^2}}
\, {\rm sin}^2 \theta &
\frac{\delta}{\sqrt{1-4\,\delta^2}} \,{\rm sin} \theta \\
-\,\frac{\delta}{\sqrt{1-4\,\delta^2}} \,{\rm cos} \theta &
-\,\frac{\delta}{\sqrt{1-4\,\delta^2}} \,{\rm sin} \theta &
\frac{1+\sqrt{1-4\,\delta^2}}{2\,\sqrt{1-4\,\delta^2}}
\end{array} \right) ~.
\nonumber
\end{eqnarray}
%
Here $\theta$ denotes the angle between the $x$~axis and the extragalactic ${\bf B}$ in a single domain.
\begin{figure}[b!]
\centering \hspace*{-6mm}
\includegraphics[width=.85\textwidth]{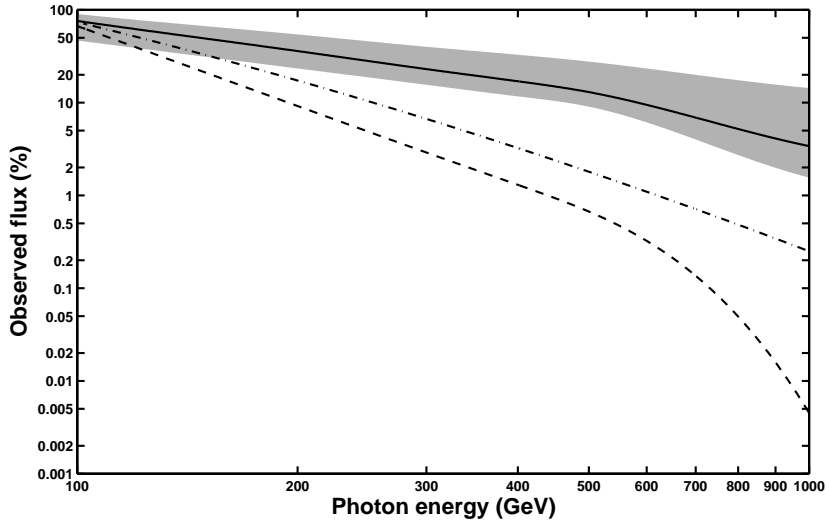}
\caption{\label{fig:comparison}
The two lowest lines give the fraction of photons surviving from a source at the same distance of 3C\,279 without the oscillation mechanism, for the ``best-fit model'' of EBL (dashed line) and for the minimum EBL density compatible with cosmology~\cite{kneiske}. The solid line represents the 
prediction of the oscillation mechanism for \mbox{$B \simeq 10^{- 9} \, {\rm G}$} and \mbox{$L_{\rm dom} \simeq 1 \, {\rm Mpc}$} within the ``best-fit model'' of EBL. The gray band is the envelope of the results obtained by independently changing ${\bf B}$ and $L_{\rm dom}$ within a factor of 10 about their preferred values.}
\end{figure}
Starting with an unpolarized photon beam, we propagate it by iterating Eq.~(\ref{eq:evolution}) as many times as the number of domains crossed by the beam, taking each time a random value for the angle $\theta$ (this reflects the random orientation of ${\bf B}$). We repeat such a procedure $10^.000$ times and finally we average over all these realizations of the propagation process. The resulting spectrum is exhibited in Fig.~\ref{fig:comparison}. We find that about 13\% of the photons arrive to the Earth for $E = 500 \, {\rm GeV}$, resulting in an enhancement by a factor of about 20 with respect to the flux expected in the absence of the proposed oscillation mechanism; the comparison is made with the ``best-fit model'' described in Kneiske {\it et al.} (2004)~\cite{kneiske}. The same calculation gives a fraction of 76\% for $E = 100 \, {\rm GeV}$ (to be compared to 67\% without the oscillation mechanism) and a fraction of 3.4\% for $E =  1 \, {\rm TeV}$ (to be compared to 0.0045\% without the oscillation mechanism). In addition, we have checked the stability of our result against independent variations of ${\bf B}$ and $L_{\rm dom}$ within a factor of 10 about their preferred values. The corresponding spectrum is represented by the gray band in Fig.~\ref{fig:comparison}. We remark that the standard deviation of the above averaging procedure lies well inside the gray band. Our prediction can be tested with the satellite-borne {\it Fermi}/LAT detector as well as with the ground-based IACTs  H.E.S.S., MAGIC, CANGAROO\,III, VERITAS and the Extensive Air Shower arrays ARGO-YBJ and MILAGRO. Our result also offers an important observational test for models of dark energy wherein quintessence is coupled to the photon through an effective dimension-five operator.

\begin{theacknowledgments}
We thank Nicola Cabibbo, Tanja Kneiske, Luciano Maiani and Massimo Persic for suggestions and comments. One of us (M. R.) thanks the Dipartimento di Fisica Nucleare e Teorica, Universit\`a di Pavia, for support. 
\end{theacknowledgments}








\bibliographystyle{aipproc}   

\bibliography{sample}

\IfFileExists{\jobname.bbl}{}
 {\typeout{}
  \typeout{******************************************}
  \typeout{** Please run "bibtex \jobname" to optain}
  \typeout{** the bibliography and then re-run LaTeX}
  \typeout{** twice to fix the references!}
  \typeout{******************************************}
  \typeout{}
 }

\end{document}